\documentclass[letterpaper, 10pt, conference]{ieeeconf}

\providecommand{\IEEEoverridecommandlockouts}{}
\providecommand{\overrideIEEEmargins}{}
\IEEEoverridecommandlockouts
\usepackage{amsmath,amssymb}
\usepackage{graphicx}
\usepackage{booktabs}
\usepackage{algorithm}
\usepackage{algorithmic}
\usepackage{xcolor}
\usepackage{url}
\usepackage{cite}
\usepackage{multirow}
\usepackage{pifont}
\usepackage{flushend}
\usepackage{hyperref}
\usepackage{cleveref}

\newcommand{\Sset}{\mathcal{S}}
\newcommand{\Gset}{\mathcal{G}}
\newcommand{\Wset}{\mathcal{W}}
\newcommand{\Xset}{\mathcal{X}}

\newcommand{\E}{\mathbb{E}}

\newcommand{\plink}{p_{\mathrm{link}}}
\newcommand{\felev}{f_{\mathrm{elev}}}

\newcommand{\Mset}{\mathcal{M}}           
\newcommand{\mok}{m_{\mathrm{OK}}}        
\newcommand{\mdead}{m^{\dagger}}          
\newcommand{\bmu}{\boldsymbol{\mu}}       
\newcommand{\ttd}{X}                      
\newcommand{\rtime}{Y}                    
\newcommand{\leth}{\ell}                  

\title{\LARGE \bf Fault-Aware MPC for \\ Robotic Fleet Communications Scheduling}

\author{Carlo Schreiber$^{1}$, Duncan Eddy$^{2}$, and Mykel J. Kochenderfer$^{2}$%
\thanks{$^{1}$Dept. of Electrical Engineering, Stanford University, USA.
$^{2}$Dept. of Aeronautics and Astronautics, Stanford University, USA.
{\tt\small \{carlops, deddy, mykel\}@stanford.edu}}}

\begin{document}

\maketitle
\thispagestyle{empty}
\pagestyle{empty}

\begin{abstract}

Operating a fleet of remote robotic systems with intermittent communications requires scheduling limited contact opportunities to maintain fleet health awareness, complete mission objectives, and intervene on faulted assets before their permanent loss. This scheduling problem is complicated by observational ambiguity: when an asset fails to check in, the operator cannot distinguish between a lethal hardware fault and a benign communications failure. If the system's failure modes are structured through a fault model, a scheduler can exploit mode-specific lethality, timing, and recoverability properties to prioritize correctly---but only if it can distinguish between modes that produce identical observations under standard actions. We present Interacting Multiple Model Model Predictive Control (IMM-MPC), a receding-horizon framework that maintains a probabilistic belief over discrete fault modes with time-inhomogeneous dynamics and optimizes a two-term objective coupling acquisition value with information gain. We characterize when observationally aliased fault modes can be disambiguated through scheduled actions and when aliasing is permanently unresolvable. Applied to satellite launch and early orbit communications scheduling, IMM-MPC recovers 59.8\% of spacecraft experiencing lethal-faults versus 9.0\% for binary-MPC and 2.0\% for a bipartite graph-based formulation solved through matching. These results hold across 200 randomized trials, while maintaining identical acquisition of healthy satellites and near-identical solve times.
\end{abstract}

\section{Introduction}
\label{sec:intro}

Remote robotic systems such as aerial drones, underwater vehicles, satellites, and planetary rovers share a common operational constraint: the operator can only communicate with each asset during brief, intermittent contact windows. Between contacts, the operator has no direct knowledge of the asset's health. A submersible vehicle surfaces on schedule for a satellite relay windows; a constellation of imaging satellites passes over a sparse network of ground stations; a fleet of long-endurance UAVs returns to radio range of a mobile command post. In each case, the operator faces the same scheduling problem: allocating limited communication opportunities across the fleet to simultaneously maintain situational awareness of fleet health, update planned mission tasks, and intervene on assets experiencing faults before those faults cause permanent loss.

The most consequential scheduling decisions occur when faults are present, not when the fleet is healthy. A satellite with a lethal power anomaly and hours to live competes for the same communications slot as a healthy satellite sending routine telemetry. The scheduler must decide which asset to contact and then must decide how to update subsequent communications plans based on the information discovered. This decision is complicated by two sources of uncertainty that interact. First, a missed contact is ambiguous: it could indicate a fatal hardware failure, a non-lethal transient, or simply an unfavorable link geometry. Second, different fault modes have different consequences---some are lethal without timely intervention, others are recoverable at leisure---but distinct modes can produce identical observations under routine operations. An unplanned link obstruction and an antenna deployment failure both present as a missed contact, yet one threatens survival and the other does not. Without a mechanism to distinguish these cases, a scheduler has no basis to prioritize the at-risk asset over one that is safe but temporarily unreachable.

Two inadequate responses dominate current practice. Binary health models collapse fault structure into a scalar alive/dead probability, treating all contact failures as equivalent evidence of degradation and discarding exactly the information needed to prioritize correctly. Deterministic fault detection waits for a confident diagnosis before acting, potentially incurring irreversible damage during the detection period~\cite{zhang2001integrated}. Neither provides a way to actively schedule actions that resolve ambiguity between fault modes while time remains for intervention.

If the system's failure modes are structured through a fault model---where each mode carries known lethality, time-to-failure, and recoverability properties---a scheduler can exploit this structure to make better allocation decisions. However, doing so requires the ability to distinguish between modes that are \emph{observationally aliased}: modes that produce identical observations under standard actions, leaving the scheduler unable to differentiate them through routine operations alone. Correct prioritization requires information-seeking actions, actions chosen specifically to disambiguate the underlying fault mode, be scheduled before irreversible failure occurs.

We propose Interacting Multiple Model Model Predictive Control (IMM-MPC), a receding-horizon framework that maintains a probabilistic belief over named discrete fault modes with time-inhomogeneous dynamics and optimizes a two-term objective coupling acquisition value with information gain. By closing the loop between belief state and scheduling decisions, IMM-MPC enables the controller to actively reduce fault-mode uncertainty and act on the resulting information before lethal faults become unrecoverable. We characterize when observationally aliased fault modes can be disambiguated through scheduled actions and when aliasing is permanently unresolvable.

We instantiate IMM-MPC on satellite Launch and Early Orbit Phase (LEOP) contact scheduling, where the problem is acute. The hours following separation from the launch vehicle combine maximum health uncertainty, resource-constrained ground station access, and likelihood of fault modes that require rapid intervention~\cite{marsh2020psyche,ksat2017leop}. However, the framework applies to any domain where an operator must decide on scheduling intermittent contacts with remote assets whose fault modes are partially observable~\cite{kochenderfer2015pomdp} and carry differential urgency.

The Interacting Multiple Model (IMM) algorithm~\cite{blom1988imm,mazor1998imm} is a Bayesian filter over finite mode sets, but prior work combining IMM with control either uses time-homogeneous transitions with no information-gathering objective~\cite{hill2021explicit} or reacts to the estimated mode without optimizing actions over the associated belief~\cite{zhang2001integrated}. Active fault diagnosis in MPC~\cite{heirung2019mpc,ferranti2019fault} introduces information-seeking costs but addresses single systems with two-mode faults and time-homogeneous transitions, without fleet scheduling or observational aliasing. Unified health management and decision-making frameworks~\cite{balaban2019unifying} and maintenance POMDP models~\cite{morato2022optimal,rokhforoz2021distributed} handle multi-state degradation with partial observability but rely on offline solvers and continuous degradation states rather than discrete fault modes. Spacecraft LEOP fault protection~\cite{marsh2020psyche,stockner2025fault} is rule-based without probabilistic multi-mode belief. IMM-MPC combines time-inhomogeneous multi-mode belief, an information-seeking scheduling objective, and a tractable scenario-based MILP into a unified receding-horizon framework that addresses all these gaps simultaneously.

The main contributions of this paper are 1) a belief-space MPC formulation with discrete fault modes and time-inhomogeneous dynamics, 2) a characterization of when aliased modes can be disambiguated, and 3) application to LEOP scheduling showing up to 8$\times$ improvement in lethal recovery.


\section{System Model and Belief Update}
\label{sec:model}

\subsection{Fault Mode Model}

Consider a fleet of $n$ assets indexed by
$i \in \{1,\ldots,n\}$, where each asset communicates with
the operator only during scheduled contact windows. Between
contacts, the operator must reason about each asset's health
using a structured fault model. At time $t_k$, each asset occupies
a discrete fault mode $m_k^i \in \Mset = \{\mok, m_1, \ldots, m_F, \mdead\}$,
where $\mok$ is nominal, $m_1,\ldots,m_F$ are distinct fault modes  (each either recoverable or
lethal), and $\mdead$ is the absorbing failure state representing permanent loss. The fault mode set and its structure are derived from system-level failure analysis. For a spacecraft this might come from a failure mode and effects analysis (FMEA).

Each fault mode $m_f$ carries three properties that determine
how the scheduler should treat it: a lethality flag $\leth_f \in \{0,1\}$ indicating whether the mode leads to
permanent failure without intervention, a time-to-death
$\ttd_f \sim p_f$ giving the elapsed time from fault onset to
$\mdead$ without intervention (defined only for lethal modes),
and a minimum recovery time $\rtime_f \sim q_f$ specifying
how long the asset must be under active intervention before it
can return to $\mok$. These three properties encode the
scheduling-relevant differences between fault modes: a lethal
mode with a short time-to-death demands immediate attention,
while a non-lethal mode with a long recovery time can be
deferred. 

Each asset's fault mode evolves according to a
row-stochastic transition matrix $\Pi(\tau_k^i)$ that depends on
the elapsed time $\tau_k^i \geq 0$ since asset $i$ entered
its current mode
\begin{equation}
  P\!\bigl(m_{k+1}^i = m' \mid m_k^i = m,\, \tau_k^i\bigr)
  = \Pi_{mm'}(\tau_k^i).
  \label{eq:transition}
\end{equation}
The dependence on elapsed time $\tau$ is what makes the dynamics time-inhomogenous: the probability of transitioning to $\mdead$ is not constant but increases the longer a lethal fault persists without intervention. Three structural requirements constrain $\Pi(\tau)$:
$\Pi_{\mdead,\mdead}(\tau) = 1$ (the dead state is absorbing---once an asset is lost, it stays lost);
$\frac{\partial}{\partial \tau}\Pi_{m_f,\mdead}(\tau) \geq 0$ for $\leth_f = 1$
(lethality monotonicity---death risk grows with elapsed time);
and $\Pi_{m_f,\mok}(\tau) = 0$ for
$\tau < \inf\,\mathrm{supp}(q_f)$
(recovery gating---an asset cannot recover before the minimum intervention time has elapsed).

For lethal modes, we parameterize the death transition using
the discrete-time hazard of $\ttd_f$
\begin{equation}
  \Pi_{m_f,\mdead}(\tau) = \alpha_f \cdot
    \frac{F_{\ttd_f}(\tau{+}\Delta t) - F_{\ttd_f}(\tau)}
         {1 - F_{\ttd_f}(\tau)},
  \label{eq:deathparam}
\end{equation}
where $F_{\ttd_f}$ is the CDF of $\ttd_f$, $\alpha_f \in (0,1)$
scales the maximum per-step probability, and $\Delta t$ is the
replanning interval. Lethality monotonicity holds for increasing failure rate (IFR)
distributions, including the uniform distributions used in~\Cref{sec:leop}~\cite{nobile2022prendiville,jang2019time}.



The controller selects action $a \in \mathcal{A}$ for asset
$i$, generating observation $o \in \mathcal{O}$ with
mode-dependent likelihood $P(o \mid m, a) = L_{m,a}(o)$.
Modes $m$ and $m'$ are observationally aliased under $a$
if $L_{m,a}(o) = L_{m',a}(o)$ for all $o$; regardless of the observation outcome, action $a$ cannot change the operator's relative belief between $m$ and $m'$. This condition holds exactly when the action has no discriminative power between the two modes under any possible outcome. A \emph{disambiguating action} $a*$ satisfies
$L_{m,a^*}(o) \neq L_{m',a^*}(o)$ for some $o$; the
information term in the IMM-MPC objective
(\Cref{sec:scheduling}) incentivizes scheduling such
actions when belief mass is divided between aliased modes.

\subsection{Belief Update}

For each asset $i$, the controller maintains a belief
vector $\bmu_k^i = [\mu_k^i(m)]_{m \in \Mset} \in
\Delta^{|\Mset|}$, where $\Delta^{|\Mset|}$ is the probability
simplex over fault modes, initialized from a prior $\bmu_0^i$ derived
from historical fault data or failure
analysis. The belief is
propagated using the IMM
filter~\cite{blom1988imm,mazor1998imm} in two steps. First,
the predicted belief accounts for mode transitions over one time step
\begin{equation}
  \bar{\mu}_{k+1}^i(m') = \sum_{m \in \Mset}
    \Pi_{mm'}(\tau_k^i)\,\mu_k^i(m),
  \quad m' \in \Mset.
  \label{eq:predict}
\end{equation}
If no action is taken on asset $i$, only this prediction
step advances $\bmu^i$. For a lethal fault, this means the belief mass drains steadily towards $\mdead$ with each step that passes without contact and the scheduler sees the asset as less likely to be recovered over time.

After executing action $a$ and observing $o$, the belief is updated using Bayes' rule
\begin{equation}
  \mu_{k+1}^i(m') = \frac{L_{m',a}(o)\,\bar{\mu}_{k+1}^i(m')}
    {\sum_{m''} L_{m'',a}(o)\,\bar{\mu}_{k+1}^i(m'')},
  \quad m' \in \Mset.
  \label{eq:update}
\end{equation}
The elapsed time $\tau_k^i$ increments by $\Delta t_k$ at each
step. For assets confirmed nominal at $t_0$, $\tau_0^i = 0$.
For those with non-negligible fault belief, the fault may have
onset before planning began, so we initialize
\begin{equation}
  \tau_0^i = \max\!\bigl(0,\; \Delta t_{\mathrm{op}} -
  \tilde{t}_{\mathrm{onset}}^i\bigr),
  \quad \tilde{t}_{\mathrm{onset}}^i \sim p_{\mathrm{onset}}^i,
  \label{eq:tau_init}
\end{equation}
where $\Delta t_{\mathrm{op}}$ is elapsed time since system
activation and $p_{\mathrm{onset}}^i$ is a prior over fault
onset derivable from historical data or failure analysis. Once initialized, $\tau_k^i$ resets only upon confirmed recovery to $\mok$.

\section{Receding-Horizon Scheduling}
\label{sec:scheduling}

The belief model in Section~\ref{sec:model} tells the operator
what it knows about each asset's fault mode; the scheduling
formulation in this section decides what to do with that
knowledge. At each replanning time $t_k$, the controller selects a schedule $x \in \Xset_k$ comprised of binary assignments $x_w \in \{0,1\}$ for each
candidate intervention $(i_w, a_w, t_w)$ in the horizon
window set $w \in \Wset_k$. Each window pairs an asset $i_w$ with an action $a_w$ at
time $t_w$, for a specific communications window $w$---concretely, this is a communication slot assigned
to a specific asset with a specific action type. Let $q_i > 0$ be the priority weight of
asset $i$ and $\theta \in (0,1)$ the recovery confidence
threshold.

\subsection{Objective Formulation}
\label{sec:objective}

The scheduler must balance two goals: 1) communicating assets to complete nominal mission
objectives and confirm fleet health and 2) allocating actions that resolve uncertainty
between aliased fault modes before lethal faults become
unrecoverable. The IMM-MPC objective couples expected acquisition value with expected information gain
\begin{equation}
  \max_{x \in \Xset_k}\;
    \E_{\bmu_k}[V(x)] + \lambda\,\hat{I}(x,\bmu_k),
  \label{eq:obj}
\end{equation}
where $\lambda \geq 0$ controls the tradeoff between the two terms.

The value term captures the expected benefit of each schedule
under the current multi-mode belief. An action on an asset in
a lethal fault mode may succeed if the mode is
responsive to the chosen action type, but the probability of
success depends on which mode is actually active
\begin{equation}
  \E[V(x)] = \sum_{i=1}^{n} q_i
    \sum_{m \in \Mset \setminus \{\mdead\}}
    \mu_k^i(m)\,P(\mathrm{success} \mid m, x^i),
  \label{eq:value}
\end{equation}
where $x^i$ is the sub-schedule for asset $i$ and success
probability is mode-dependent through~\eqref{eq:update}. This value term incorporates the belief state: an asset with high $\mu(\mdead)$ contributes little expected value regardless of how many windows are allocated to it, so the scheduler naturally avoids using windows on likely-dead assets in favor of ones that can still be saved.

The information term incentivizes active disambiguation of aliased
fault pairs
\begin{equation}
  I(x,\bmu_k) = \sum_{i=1}^{n}
    \bigl[H(\bmu_k^i) -
    \E_{o \mid x^i,\bmu_k^i}[H(\bmu_{k+1}^i)]\bigr],
  \label{eq:info}
\end{equation}
where $H(\bmu) = -\sum_m \mu(m)\log\mu(m)$ is the entropy of the belief. This is the expected reduction in uncertainty about asset $i$'s fault mode from executing actions assigned to it. Since $I$ is nonlinear in $x$, we precompute a per-window linearization. For each candidate window $w$, we compute the expected entropy reduction that scheduling $w$ alone would provide
\begin{equation}
  \Delta H_w := H(\bmu_k^{i_w}) -
    \sum_{o \in \mathcal{O}}
      P(o \mid \bmu_k^{i_w}, a_w)\,
      H\!\bigl(\bmu_{k+1}^{i_w,o}\bigr),
  \label{eq:deltaH}
\end{equation}
giving $\hat{I}(x,\bmu_k) = \sum_{w \in \Wset_k} \Delta H_wx_w$, linear in $x$ and precomputable from $\bmu_k$.

$\Delta H_w > 0$ only when
$a_w$ disambiguates an aliased pair with non-negligible belief
mass. If all modes with positive belief mass produce identical observation distributions under $a_w$, then $\Delta H_w = 0$ and the information term contributes nothing for that window. In practice $\hat{I}$ is sparse: most windows use standard actions that do not disambiguate and only windows with disambiguating actions on assets with split belief receive nonzero information value. Urgency is handled
implicitly through the transition dynamics rather than as a separate objective
term.

\subsection{Scenario Approximation and MILP}
\label{sec:milp}

The expected value term~\eqref{eq:value} requires integrating over the joint distribution of fault modes across the fleet and the planning horizon. We approximate $\E[V(x)]$ through Monte Carlo sampling over mode trajectories~\cite{kleywegt2002saa,birge2011stochastic}. For
each scenario $s \in \{1,\ldots,S\}$ and asset $i$, a
correlated mode trajectory is drawn from the current belief
\begin{equation}
  \begin{split}
  \tilde{m}_k^{i,(s)} &\sim \bmu_k^i, \\
  \tilde{m}_{k+t}^{i,(s)} \mid \tilde{m}_{k+t-1}^{i,(s)}
    &\sim \Pi_{\tilde{m}_{k+t-1}^{i,(s)},\cdot}
    \!\bigl(\tau_k^i + t\Delta t\bigr),
    \quad t = 1,\ldots,H.
  \end{split}
  \label{eq:scengen}
\end{equation}
The initial mode is sampled from the current belief and subsequent modes evolve according to the time-inhomogenous transition matrix. This produces a complete fault-trajectory for each asset in the scenario, capturing the correlation between windows: if asset $i$ transitions to $\mdead$ at any point in scenario $s$ then every subsequent window for asset $i$ in that scenario will have zero probability of success.

For each window $w$ in scenario $s$, a binary outcome is
sampled
\begin{equation}
  \begin{split}
  z_{w,s} &:=
    \mathbf{1}\!\bigl[\tilde{m}_{t_w}^{i_w,(s)}
      \in \Mset_{\mathrm{resp}}\bigr]
    \cdot Z_{w,s}, \\
  Z_{w,s} &\sim
    \mathrm{Bernoulli}\!\bigl(
      p_{\mathrm{succ}}(w,\tilde{m}_{t_w}^{i_w,(s)})\bigr),
  \end{split}
  \label{eq:zws}
\end{equation}
where $\Mset_{\mathrm{resp}} \subseteq \Mset$ is the set of
modes that can respond to action $a_w$. The shared trajectory
across all windows of asset $i$ encodes joint fault
correlation: if $\tilde{m}^{i,(s)} = \mdead$, then $z_{w,s} = 0$
for every window of that asset in scenario $s$. Since every
$z_{w,s}$ is a fixed precomputed coefficient, the problem
reduces to a single deterministic MILP
\begin{align}
  \max_{x,\,a}\;\; &
    \frac{1}{S}\sum_{s=1}^{S}\sum_{i=1}^{n} q_i\,a_{i,s}
    + \lambda\,\hat{I}(x,\bmu_k)
  \label{eq:milp_obj} \\
  \text{s.t.}\quad
    & a_{i,s} \leq \sum_{w:\,i_w=i} x_w\,z_{w,s},
      \quad \forall i,s,
    \label{eq:milp_acq} \\
    & x \in \Xset_k,\quad
      a_{i,s} \in [0,1],\quad
      x_w \in \{0,1\},
    \label{eq:milp_bin}
\end{align}
with $O(n \cdot S \cdot |\Wset_k|)$ variables~\footnote{the scenario approximation converges to the true optimum for the 
continuous relaxation~\cite{kleywegt2002saa}; $S = 50$ suffices in practice.}. The auxiliary variable $a_{i,s}$ captures whether asset $i$ is successfully acquired in scenario $s$; constraint~\eqref{eq:milp_acq} ensures this can only happen if at least one scheduled window for asset $i$ succeeds in that window.

At each replanning step, scenarios are drawn from $\bmu_k$, the MILP
is solved, the earliest selected action
$w^\star = \arg\min_{w:\,x_w^\star=1} t_w$ is executed, and
the belief updates via~\eqref{eq:predict}--\eqref{eq:update}.
An asset is declared acquired when
$\mu_{k+1}^{i^\star}(\mok) > \theta$. The controller then replans from the updated belief, implemented only the first action of each plan in the standard receding-horizon fashion.


\section{Satellite LEOP Scheduling Problem}
\label{sec:leop}

The framework described in Sections~\ref{sec:model}
and~\ref{sec:scheduling} applies to any fleet scheduling
problem with intermittent contacts, structured fault modes,
and partial observability. We instantiate it on a domain
where all of these features are present in an operationally
demanding form, contact scheduling during the launch and early operations phase
of a space mission.

LEOP is the period immediately following separation from the
launch vehicle. The operator must establish contact with each
satellite, confirm nominal health, and intervene on any faults
before they cause permanent loss. The problem is challenging for
three reasons. First, health uncertainty is at its maximum:
every satellite has just experienced the mechanical and thermal
stresses of launch, and the operator has no telemetry until
the first successful contact. Second, communication resources
are severely constrained: ground station networks provide only
brief contact windows as satellites pass overhead, and
multiple satellites compete for the same station slots. Third,
several fault modes are lethal on timescales of hours---a
satellite with a deployment anomaly drains its batteries and
dies if not reached in time. This combination of high
uncertainty, scarce resources, and time-critical faults makes
LEOP a natural benchmark for fault-mode-aware scheduling.

\subsection{Assets, Actions, and Contact Windows}

Assets are satellites $\Sset = \{1,\ldots,n_s\}$ and
communication opportunities are ground contact windows between
satellite $s_w \in \Sset$ and ground station $g_w \in \Gset$,
defined by the tuple $w = (s_w,\, g_w,\, t_w^{\mathrm{start}},\,
t_w^{\mathrm{end}},\, e_w^{\max},\, a_w)$, computed from TLE-based SGP4 propagation at a $5^\circ$
elevation mask. Each window carries an action type
$a_w \in \{a_{\mathrm{contact}},\, a_{\mathrm{beacon}}\}$:
full S-band frequency contacts attempt two-way telemetry and command
upload, while UHF frequency beacon listens passively detect the
satellite's emergency beacon. Both consume one ground station
slot and compete equally in the MILP.

\subsection{Fault Modes}

The fault mode set $\Mset$, comprising the five modes summarized 
in Table~\ref{tab:modes}, reflects the principal early-orbit 
failure categories~\cite{marsh2020psyche,ksat2017leop}. Two modes 
are lethal: \textsc{GNC} (attitude or navigation fault) and \textsc{DEP} (deployment
anomaly), both draining batteries within 6--24 hours without
intervention. \textsc{COMMS} (transceiver anomaly) is non-lethal---the
satellite is healthy but cannot communicate through the
primary S-band link. \textsc{DEAD} is the absorbing failure state.

\begin{table}[t]
  \centering
  \caption{LEOP fault mode characteristics.}
  \label{tab:modes}
  \setlength{\tabcolsep}{3.5pt}
  \small
  \begin{tabular}{@{}lcccc@{}}
    \toprule
    Mode & $\leth$ & $\ttd$ (hr) & $\rtime$ (min) & Description \\
    \midrule
    \textsc{OK}    & 0 & $\times$            & $\times$
          & Nominal \\
    \textsc{GNC}   & 1 & $\mathrm{U}(6,24)$ & $\mathrm{U}(15,120)$
          & Attitude/nav fault \\
    \textsc{COMMS} & 0 & $\times$            & $\mathrm{U}(15,120)$
          & Transceiver anomaly \\
    \textsc{DEP}   & 1 & $\mathrm{U}(6,24)$ & $\mathrm{U}(15,120)$
          & Deployment anomaly \\
    \textsc{DEAD}  & -- & --                 & --
          & Absorbing failure \\
    \bottomrule
  \end{tabular}
\end{table}

\textsc{GNC} faults are immediately distinguishable on successful
contact: the satellite returns an attitude error flag in its
telemetry, so a single successful contact resolves the mode.
The scheduling challenge for \textsc{GNC} is reaching the satellite
before it dies, not identifying the fault. \textsc{COMMS} and \textsc{DEP} are the aliased pair. Under
$a_{\mathrm{contact}}$, both produce identical observations:
$P(\mathrm{CONTACT} \mid \textsc{COMMS}) =
P(\mathrm{CONTACT} \mid \textsc{DEP}) = 0$. A \textsc{COMMS} fault
disables the S-band transmitter, and a \textsc{DEP} fault results in
the obstruction of both S-band and UHF transmitting antennas, preventing their
use---in both cases, the contact attempt fails. 
No sequence of full contacts can
change the ratio $\frac{\mu(\textsc{COMMS})}{\mu(\textsc{DEP})}$.
Every failed contact reinforces the belief that something is
wrong, but provides no information about whether the fault is
lethal.

Under $a_{\mathrm{beacon}}$, the two modes separate. A COMMS
fault leaves the UHF emergency beacon functional---the
satellite is healthy and powered, it simply cannot communicate
on S-band---so $P(\mathrm{BEACON} \mid \textsc{COMMS}) = p_b$.
A DEP fault cuts power to all RF systems, so
$P(\mathrm{BEACON} \mid \textsc{DEP}) = \epsilon \approx 0$.
Listening for beacon pings are therefore disambiguating actions:
scheduling a beacon listen on
a satellite with split \textsc{COMMS}/\textsc{DEP} belief produces
$\Delta H_w > 0$, and the information term incentivizes
allocating these slots before \textsc{DEP}'s lethality clock expires.

\subsection{Observation and Link Models}

Contact success probability follows a logit model that
depends on the maximum elevation angle of the pass
\begin{equation}
  \plink(w) = \sigma\!\bigl(
    \beta_0 + \beta_e \felev(e_w^{\max})\bigr),
  \label{eq:plink}
\end{equation}
with $\felev(e) = \log(\sin(\max(e,e_{\mathrm{floor}})\pi/180))$.
Higher elevation passes close the link more reliably, which
the scheduler accounts for when choosing between competing
windows. The full observation likelihood table is given in
Table~\ref{tab:obs}, where $p_b \in (0,1)$ is the beacon
response probability for RF-functional modes, $\gamma$ is the
contact success degradation factor for \textsc{GNC} faults (the
satellite is tumbling, reducing antenna gain), and
$\epsilon \approx 0$ reflects near-certain RF failure under a
deployment anomaly.

\begin{table}[t]
  \centering
  \caption{Observation likelihoods by mode and action type.}
  \label{tab:obs}
  \setlength{\tabcolsep}{3pt}
  \small
  \begin{tabular}{@{}lccccc@{}}
    \toprule
    & \multicolumn{2}{c}{$a_{\mathrm{contact}}$}
    & $a_{\mathrm{beacon}}$ \\
    \cmidrule(lr){2-3} \cmidrule(lr){4-4}
    Mode & $P(\mathrm{CONT.})$ & $P(\mathrm{NO\_CONT.})$
         & $P(\mathrm{BEACON})$ \\
    \midrule
    \textsc{OK}    & $\plink$              & $1{-}\plink$      & $p_b$ \\
    \textsc{GNC}   & $\plink{\cdot}\gamma$ & $1{-}\plink{\cdot}\gamma$
          & $p_b$ \\
    \textsc{COMMS} & $0$ & $1$ & $p_b$ \\
    \textsc{DEP}   & $0$ & $1$ & $\epsilon$ \\
    \textsc{DEAD}  & $0$ & $1$ & $0$ \\
    \bottomrule
  \end{tabular}
\end{table}

\section{Experiments}
\label{sec:experiments}

We evaluate IMM-MPC on two LEOP scenarios that differ in
ground station coverage and fleet size. The first uses a
well-resourced global network where the scheduler has enough
windows to allocate disambiguating actions alongside routine
contacts. The second uses a sparse regional network where
every misallocated slot carries high opportunity cost. Together,
the two scenarios test whether the information term provides
value across different resource regimes.

All scenarios implemented the Brahe astrodynamics toolbox~\cite{eddy2026brahemodernastrodynamicslibrary} and the
logit contact model~\eqref{eq:plink} with $\beta_0 = 0.8$,
$\beta_e = 1.5$, and $\gamma = 0.4$, and beacon listens with
$p_b = 0.8$ and $\epsilon = 0.05$. IMM-MPC uses $S = 50$
stratified scenarios, horizon $H = 3$\,hr, $\lambda = 0.1$,
and Gurobi~10+ with a 10\,s solve limit over 200 Monte Carlo
trials. At separation, 70\% of satellites are nominal, 15\%
have already failed, and 15\% carry active faults drawn from 
$\{\textsc{GNC}, \textsc{COMMS}, \textsc{DEP}\}$ of which approximately
70\% are lethal. In the LEOP instantiation, faults are assumed to onset at
separation, so active-fault satellites are initialized with
$\tau_0^i = t_{\mathrm{sep}}$. Satellites are declared acquired when $\mu_{k+1}^{i^\star}(\mok) > 0.99$.

We compare IMM-MPC against two baselines that represent a natural progression of scheduling sophistication. Bipartite matching assigns satellites to windows greedily by
priority and elevation with no health model or lookahead: we construct a bipartite graph with 
satellites on one side and contact windows on the other, weighted 
by the product of satellite priority $q_i$ and link probability 
$\plink(w)$, and solve for a maximum-weight matching using the 
Hungarian algorithm, 
representing current operational practice in most fleet management systems. Binary-MPC replaces
the multi-mode belief with a scalar aliveness probability per
satellite, optimizing a single-scenario MILP with no information term. This adds probabilistic health reasoning but
collapses all fault modes into a single dimension. IMM-MPC
maintains the full multi-mode belief and optimizes the
two-term objective from Section~\ref{sec:scheduling}.

\subsection{Rideshare Scenario}

The first scenario we consider is inspired by small commercial operators, launching
on SpaceX Transporter rideshare launches. We simulate 100 satellites in a sun-synchronous orbit ($\sim$525\,km, 97.4$^\circ$) served by
6 KSAT global ground stations over 48 hours\footnote{SvalSat 
(Svalbard, Norway), TrollSat (Antarctica), Punta Arenas 
(Argentina), Hartebeesthoek (South Africa), Awarua (New Zealand), 
and Athens (Greece)~\cite{eddy2025ground}.}, with prior
$\mu_0(\mok) = 0.90$. Most satellites have 3--6 contact windows in the horizon, giving the information term multiple opportunities
to take beacon contacts for \textsc{COMMS}/\textsc{DEP} disambiguation.

\begin{table}[H]
  \centering
  \caption{KSAT results (200 trials, 6 stations, 100 sats,
    48\,hr).}
  \label{tab:ksat}
  \setlength{\tabcolsep}{3pt}
  \small
  \begin{tabular}{@{}lrrr@{}}
    \toprule
    Metric & Bipartite & Binary & IMM-MPC \\
    \midrule
    Overall (\%)
      & $59.5 \pm 1.2$ & $83.4 \pm 1.0$ & $\mathbf{90.0 \pm 1.8}$ \\
    Lethal rec.\ (\%)
      & $2.0 \pm 2.1$ & $9.0 \pm 7.5$ & $\mathbf{59.8 \pm 13.8}$ \\
    Solve time (s)
      & $<\!1$ & $32.4$ & $34.5$ \\
    \bottomrule
  \end{tabular}
\end{table}

As seen in \Cref{tab:ksat}, IMM-MPC recovers 59.8\% of lethal-fault satellites versus
9.0\% for binary-MPC ($6.6\times$) and 2.0\% for bipartite
matching. The gap comes directly from disambiguation:
binary-MPC cannot distinguish \textsc{COMMS} from \textsc{DEP} faults after a failed
contact, so it treats the satellite as low-value and allocates
slots elsewhere to confirm health of other, uncontacted spacecraft. IMM-MPC recognizes the unresolvable aliasing
under $a_{\mathrm{contact}}$, schedules a beacon listen to
break it, and escalates or deprioritizes based on the result.
Solve time increases by 2.1\,s. The wide confidence interval
on lethal recovery ($\pm 13.8\%$) reflects the small number of
recoverable lethal instances per trial.

\begin{figure}[htb]
  \centering
  \includegraphics[width=\columnwidth]{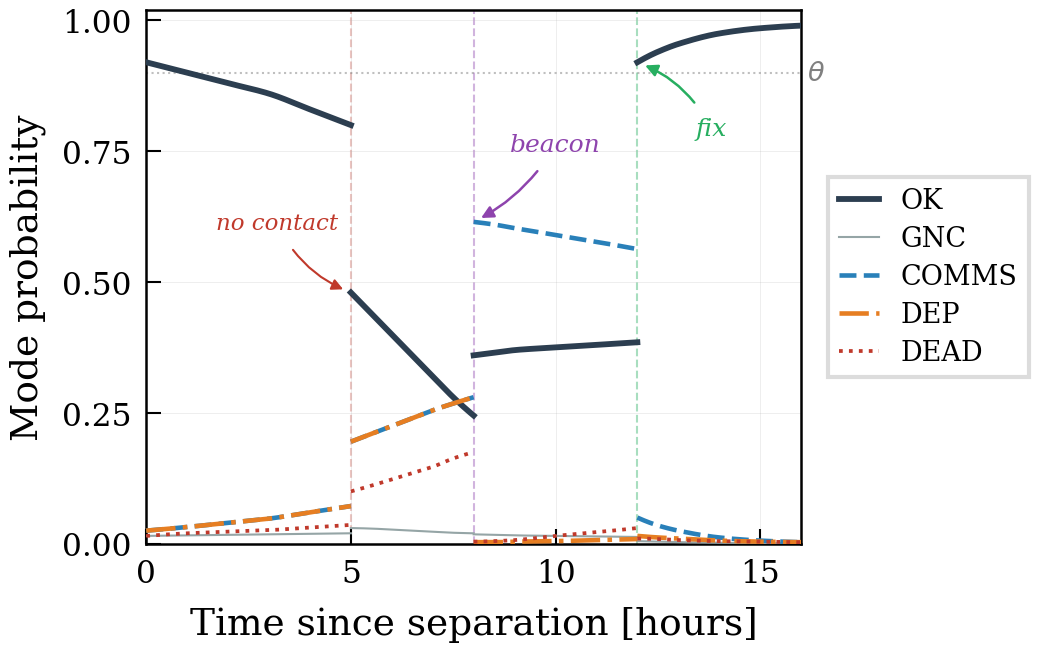}
  \caption{IMM belief trajectory for a \textsc{COMMS} fault.
  Contact failures leave $\frac{\mu(\textsc{COMMS})}{\mu(\textsc{DEP})}$
  invariant; a beacon listen
  at $t \approx 8$\,hr concentrates belief on \textsc{COMMS} over \textsc{DEP},
  enabling recovery.}
  \label{fig:belief_traj}
\end{figure}

\begin{figure}[htb]
  \centering
  \includegraphics[width=\columnwidth]{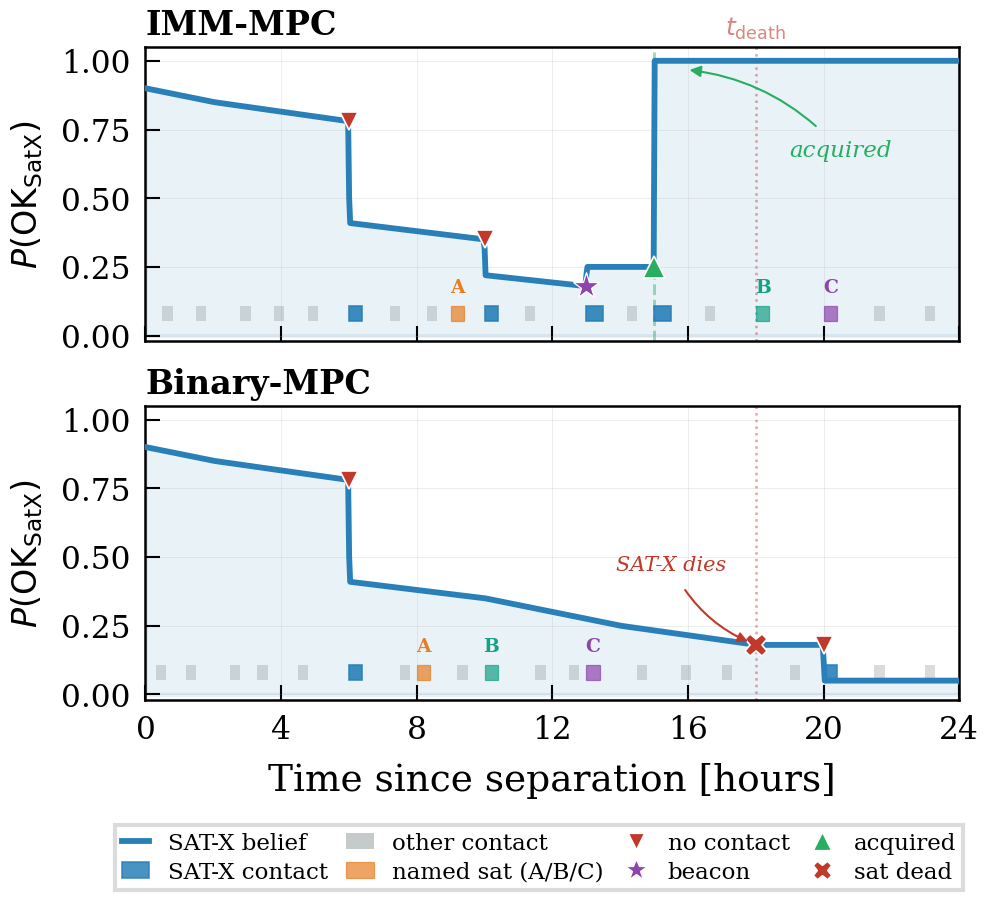}
  \caption{State-aware triage on a \textsc{DEP}-fault satellite.
  IMM-MPC recognizes lethal fault probability and prioritizes
  the at-risk satellite, achieving recovery before
  $t_{\mathrm{death}}$ (red dashed). Binary-MPC allocates
  the same slots to nominal satellites~A,~B,~C and arrives
  too late.}
  \label{fig:triage}
\end{figure}

Figure~\ref{fig:belief_traj} shows this mechanism on a single
COMMS-fault satellite: failed contacts leave the
$\frac{\mu(\textsc{COMMS})}{\mu(\textsc{DEP})}$ ratio invariant, and
a beacon listen concentrates belief on \textsc{COMMS}, freeing future
slots for higher-urgency assets.
Figure~\ref{fig:triage} shows the triage consequence: IMM-MPC
prioritizes a \textsc{DEP}-fault satellite and recovers it before
$t_{\mathrm{death}}$, while binary-MPC allocates the same
slots to nominal satellites and arrives too late.

Figure~\ref{fig:belief_traj} illustrates the belief dynamics on a single satellite with a \textsc{COMMS} fault. Each failed contact
shifts belief mass away from OK but leaves the \textsc{COMMS}/\textsc{DEP} ratio $\frac{\mu(\textsc{COMMS})}{\mu(\textsc{DEP})}$
unchanged---exactly the aliasing invariance established in
Section~\ref{sec:model}. The information term schedules a 
beacon listen, which produces a beacon detection (the UHF
system is functional under \textsc{COMMS}) and concentrates belief on
\textsc{COMMS}. With the lethal mode ruled out, the scheduler
deprioritizes this satellite and reallocates its future windows
to assets with higher urgency.

Figure~\ref{fig:triage} illustrates the state-aware triage
behavior behind the lethal recovery improvement. Both schedulers observe
identical contact failures for a satellite with a \textsc{DEP} fault, but IMM-MPC maintains explicit
belief over lethal fault modes and their lethality countdowns,
recognizes that the satellite may be
in a lethal mode with limited time remaining, and assigns it
higher priority than nominal satellites competing for the same
slots. Binary-MPC lacks any representation of mode lethality; it
sees a satellite with low aliveness probability and treats it
as a low-value target, allocating the same slots to healthy
satellites instead. By the time binary-MPC returns to the
faulted satellite, it has passed $t_{\mathrm{death}}$. The outcome difference
is structural: state-aware scheduling maintains fault mode identity
and recovers the satellite where a scalar health model cannot.

\subsection{Megaconstellation Scenario}

The second scenario tests the framework under resource
scarcity. This scenario covers 140 satellites in SSO at ${\sim}500$\,km served by 3 ground stations (Hawaii, Mauritius, Alpine Texas) over 72 hours, reflecting a resource-constrained megaconstellation launch where only a few stations can be dedicated to launch support. This scenario uses prior $\mu_0(\mok) = 0.85$.
Most satellites receive only 1--2 windows; every
misallocated slot carries high opportunity cost.

\begin{table}[htb]
  \centering
  \caption{Megaconstellation results (200 trials, 3 stations,
    140 sats, 72\,hr).}
  \label{tab:mega}
  \setlength{\tabcolsep}{3pt}
  \small
  \begin{tabular}{@{}lrrr@{}}
    \toprule
    Metric & Bipartite & Binary & IMM-MPC \\
    \midrule
    Overall (\%)
      & $41.2 \pm 1.5$ & $82.5 \pm 1.1$ & $\mathbf{85.1 \pm 1.8}$ \\
    Lethal rec.\ (\%)
      & $0.4 \pm 0.9$ & $3.0 \pm 3.2$ & $\mathbf{24.1 \pm 9.6}$ \\
    Solve time (s)
      & $<\!1$ & $31.8$ & $33.2$ \\
    \bottomrule
  \end{tabular}
\end{table}

IMM-MPC recovers 24.1\% of lethal satellites versus 3.0\%
for binary-MPC, an $8\times$ improvement. The lower absolute 
rate reflects the resource constraint: with 1--2 windows per satellite, many
lethal faults expire before any contact is possible regardless
of scheduling. The bipartite baseline (0.4\%) confirms that
current practice approaches zero lethal recovery under severe
scarcity. Overall acquisition is comparable across methods
(82.5\% binary-MPC vs.\ 85.1\% IMM-MPC), confirming that the
information term improves lethal recovery without cost to
nominal fleet throughput.

Across both scenarios, the pattern is consistent: the
information term adds value specifically where aliased modes
prevent the value term from correctly prioritizing at-risk
assets, with the benefit scaling with available coverage for
disambiguating actions.

\section{Conclusion}
\label{sec:conclusion}

Scheduling communication with a fleet of remote assets under
intermittent contact and partial observability requires
reasoning about structured fault modes, not just scalar health
estimates. We presented IMM-MPC, a receding-horizon framework
that maintains a probabilistic belief over discrete fault
modes with time-inhomogeneous dynamics, characterizes
when observationally aliased modes can be disambiguated through
scheduled actions, and optimizes a two-term objective coupling
acquisition value with information gain. On two satellite LEOP
scenarios, IMM-MPC recovers up to $8\times$ more lethal-fault
satellites than binary-MPC ($6.6\times$ rideshare, $8\times$
megaconstellation) while maintaining comparable overall
acquisition and near-identical solve times---with the
improvement coming entirely from exploiting fault mode
structure through active disambiguation. While instantiated
here on spacecraft LEOP scheduling, the formulation is general
to any domain where an operator schedules intermittent contacts
with remote assets whose fault modes are partially observable
and carry differential urgency. Future work includes a
continuous-time formulation for asynchronous contact windows,
correlated fault priors to capture common-cause failures, and
extension to non-spacecraft domains with the same
intermittent-contact scheduling structure.

\bibliographystyle{IEEEtran}
\bibliography{references}

\end{document}